\documentclass[]{spie}  %>>> use for US letter paper
\usepackage{amsmath,amsfonts,amssymb}
\usepackage{graphicx}
\usepackage[colorlinks=true, allcolors=blue]{hyperref}

\title{The Photo-z Infrared Telescope (PIRT) – a space instrument for rapid follow up of high-redshift gamma-ray bursts and electromagnetic counterparts to gravitational wave events}

\author[a]{M.~Seiffert}
\author[a]{A.~Balady}
\author[a]{T.-C.~Chang}
\author[a]{R.~Dyer}
\author[b,c]{H. Fausey}
\author[b,c]{S.~Guiriec}
\author[a]{M.~Hart}
\author[a]{R.O.~Morris}
\author[a]{J.I.~Rodriguez}
\author[d]{P.~Roming}
\author[a]{M.~Rud}
\author[a]{D.~Russell}
\author[e]{R.~Sambruna}
\author[a]{R.~Terrile}
\author[a]{V.~Torossian}
\author[b,c]{A.J.~van~der~Horst}
\author[b,c]{N.E.~White}
\author[a]{P.~Willems}
\author[a]{A.~Woodmansee}
\author[f]{E.T.~Young}

\affil[a]{Jet Propulsion Laboratory, California Institute of Technology, 4800 Oak Grove Dr., Pasadena, CA 91109, USA}
\affil[b]{Department of Physics, The George Washington University, 725 21st Street NW, Washington, DC 20052, USA}
\affil[c]{Astronomy, Physics, and Statistics Institute of Sciences (APSIS), 725 21st Street NW, Washington, DC 20052, USA}
\affil[d]{Southwest Research Institute, 6220 Culebra Rd, San Antonio, TX 78238, USA}
\affil[e]{NASA Goddard Space Flight Center, 8800 Greenbelt Rd, Greenbelt, MD 20771, USA}
\affil[f]{Universities Space Research Association, 615 National Avenue, Suite 220
Mountain View, CA 94043, USA}

\authorinfo{E-mail: michael.d.seiffert@jpl.nasa.gov}

\begin{document} 

\newcommand\aj{AJ} 
          % Astronomical Journal 
\newcommand\araa{ARA\&A} 
          % Annual Review of Astron and Astrophys 
\newcommand\apj{ApJ} 
          % Astrophysical Journal 
\newcommand\apjl{ApJ} 
          % Astrophysical Journal, Letters 
\newcommand\apjs{ApJS} 
          % Astrophysical Journal, Supplement 
\newcommand\procspie{Proc.~SPIE} 
          % Proceedings of the SPIE 
\newcommand\pasp{PASP}
          % Publications of the Astronomical Society of the Pacific
\newcommand\nat{Nature}
          % Publications of the Astronomical Society of the Pacific

\maketitle

\begin{abstract}
The Photo-z InfraRed Telescope (PIRT) is an instrument on the Gamow Explorer, currently proposed for a NASA Astrophysics Medium Explorer.  PIRT works in tandem with a companion wide-field instrument, the Lobster Eye X-ray Telescope (LEXT), that will identify x-ray transients likely to be associated with high redshift gamma-ray bursts (GRBs) or electromagnetic counterparts to gravitational wave (GW) events.  After receiving an alert trigger from LEXT, the spacecraft will slew to center the PIRT field of view on the transient source.  PIRT will then begin accumulating data simultaneously in five bands spanning 0.5 – 2.5 microns over a 10 arc-minute field of view. Each PIRT field will contain many hundreds of sources, only one of which is associated with the LEXT transient. PIRT will gather the necessary data in order to identify GRB sources with redshift $z > 6$, with an expected source localization better than 1 arcsec. A near real-time link to the ground will allow timely follow-up as a target of opportunity for large ground-based telescopes or the James Webb Space Telescope (JWST).  PIRT will also allow localization and characterization of GW event counterparts. We discuss the instrument design, the on-board data processing approach, and the expected performance of the system. 

\end{abstract}


\section{INTRODUCTION}
\label{sec:intro}  % \label{} allows reference to this section

%%%%%%%%%%%%%%%%%%%%%%%%%%%%%%%%%%%%%%%
% Gamow configuration figure indicating PIRT
   \begin{figure} [ht]
   \begin{center}
   \begin{tabular}{c} %% tabular useful for creating an array of images 
   \includegraphics[width=15cm]{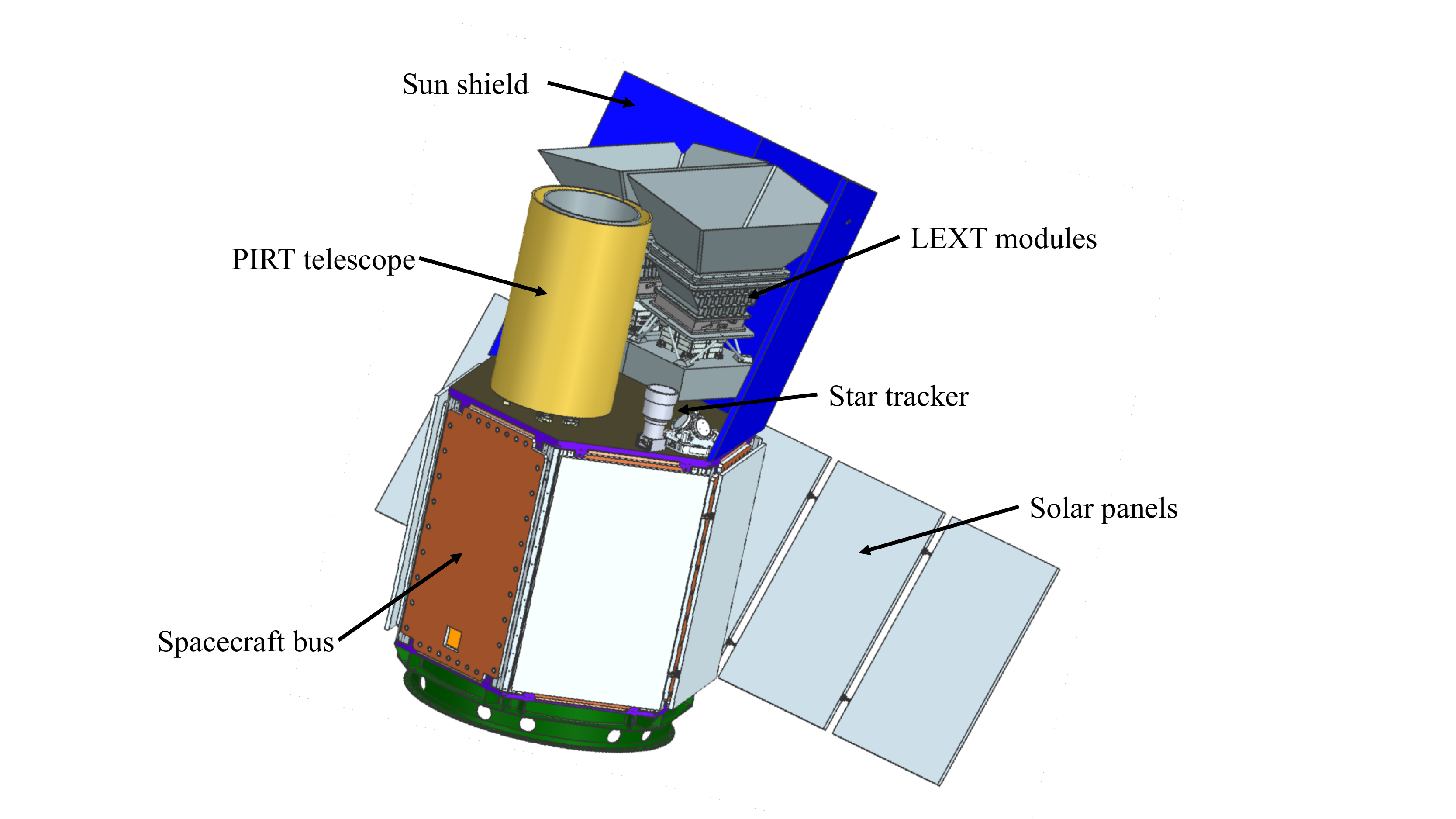}
   \end{tabular}
   \end{center}
   \caption[example] { \label{fig:Gamow} 
Overview of the Gamow flight system indicating the configuration of the instruments and their mounting on the spacecraft bus. }
   \end{figure} 
%%%%%%%%%%%%%%%%%%%%%%%%%%%%%%%%%%%%%%

We are planning to develop an optical to near-infrared instrument, the Photo-z InfraRed Telescope (PIRT), as part of the proposed Gamow Explorer MIDEX astrophysics mission.  The mission will conduct continuous science observations for a period of no less than 3 years from an orbital position at the Sun-Earth L2 point.  The PIRT instrument is intended to provide rapid follow-up observations of potential high redshift gamma-ray bursts (GRBs) as detected by the companion Lobster Eye X-ray Telescope (LEXT) instrument. PIRT's higher angular resolution and multi-band detection system allows precise position location and redshift estimation of the sources detected by LEXT. 

The PIRT instrument is designed for simplicity. It comprises an all-aluminum telescope, a compact five-band dichroic assembly, and a 2.5 micron cutoff infrared detector.  The baseline approach is to use a passively cooled focal plane and electronics assembly.  A schematic overview of the configuration of the PIRT and LEXT instruments in the Gamow mission is indicated in Figure~\ref{fig:Gamow}. 

Below we describe the scientific motivation for the mission, describe the instrument and its components, describe the necessary data processing, and give an overview of the expected performance of the system.

\section{SCIENTIFIC MOTIVATION}
\label{sec:sci}
Since their initial discovery \cite{Klebesadel73}, 
GRBs have become a unique and powerful probe of our universe. Long GRBs, defined as those with an event duration of greater than 2 seconds, can arise from the relativistic jet created by the core collapse supernova explosion of a massive star and its subsequent black hole formation \cite{Woosley2006}. These bursts are the most luminous explosions known and are observable to high redshift ($z \sim 9$ is the current record)\cite{Cucchiara2011}. Although only a few high redshift GRBs have been detected to date, the prospect of their detection in greater numbers, along with precise position determination and redshift measurement, raises the possibility of using them as a cosmological probe.  One of the most exciting prospects is the use of GRB afterglows as a ``backlight'' to probe the reionization history of the universe  \cite{Lidz2021}. 

Short GRBs are defined as bursts with duration of less than 2 seconds. One short GRB has been associated with a gravitational wave (GW) event in which a jet emerged from the black hole formed in a binary neutron star (BNS) merger \cite{Abbott2017}.
Although the rate of BNS mergers remains uncertain, future advances in ground-based gravitational wave observatories are expected to dramatically increase their detection rate.  The LEXT and PIRT instruments will allow identification of the electromagnetic counterparts to such events and thereby enable a wealth of unique studies, including jet physics, tests of quantum gravity, and r-process heavy element formation \cite{Kasen2017}.

The proposed Gamow Explorer mission has been conceived as an observatory to detect, determine precise positions, determine approximate redshifts, and ultimately coordinate with other observatories to measure spectra of a collection of $z>6$ GRBs and to find and study a collection of counterparts to gravitational wave events. The science objectives can be summarized as: 
\begin{itemize}
    \item Map the reionization history of the intergalactic medium (IGM)
    \item Trace the chemical  enrichment history at high redshift
    \item Survey GRB production at high redshift ($z > 6$), and
    \item Search for X-ray and optical-IR counterparts for GW events and rapidly provide their location to other telescopes for follow up observations.
\end{itemize}
In addition to the primary science objectives, Gamow's instruments will enable a variety of time-domain astrophysics studies that can be conducted in parallel. PIRT will conduct a galaxy survey to identify infrared transients during the course of its primary task of waiting for GRB or GW events. The galaxy survey will be based on the transient survey pioneered by the Spitzer Infra-Red Intensive Transient Survey (SPIRITS)\cite{Kasliwal2017}.
The Gamow science objectives are described in more detail in a companion paper \cite{White2021}.

\section{INSTRUMENT DESCRIPTION}
\label{sec:instrument}
The principal task for the PIRT instrument is to respond to triggers from the companion LEXT instrument.  LEXT has a relatively large field-of-view, of order 1350~deg$^2$, and a position resolution of order 2 arcmin. After the spacecraft slews to center PIRT on the LEXT source position, PIRT will gather photometric data that will allow precise position location, of order 1 arcsec, and a photometric redshift determination.  The higher location precision and the redshift estimation are critical pieces of information that will be conveyed to other observatories for rapid follow-up to produce deep, moderate resolution spectra needed for the scientific goals cited earlier. 

\begin{figure} [ht]
   \begin{center}
   \begin{tabular}{c} %% tabular useful for creating an array of images 
   \includegraphics[width=15cm]{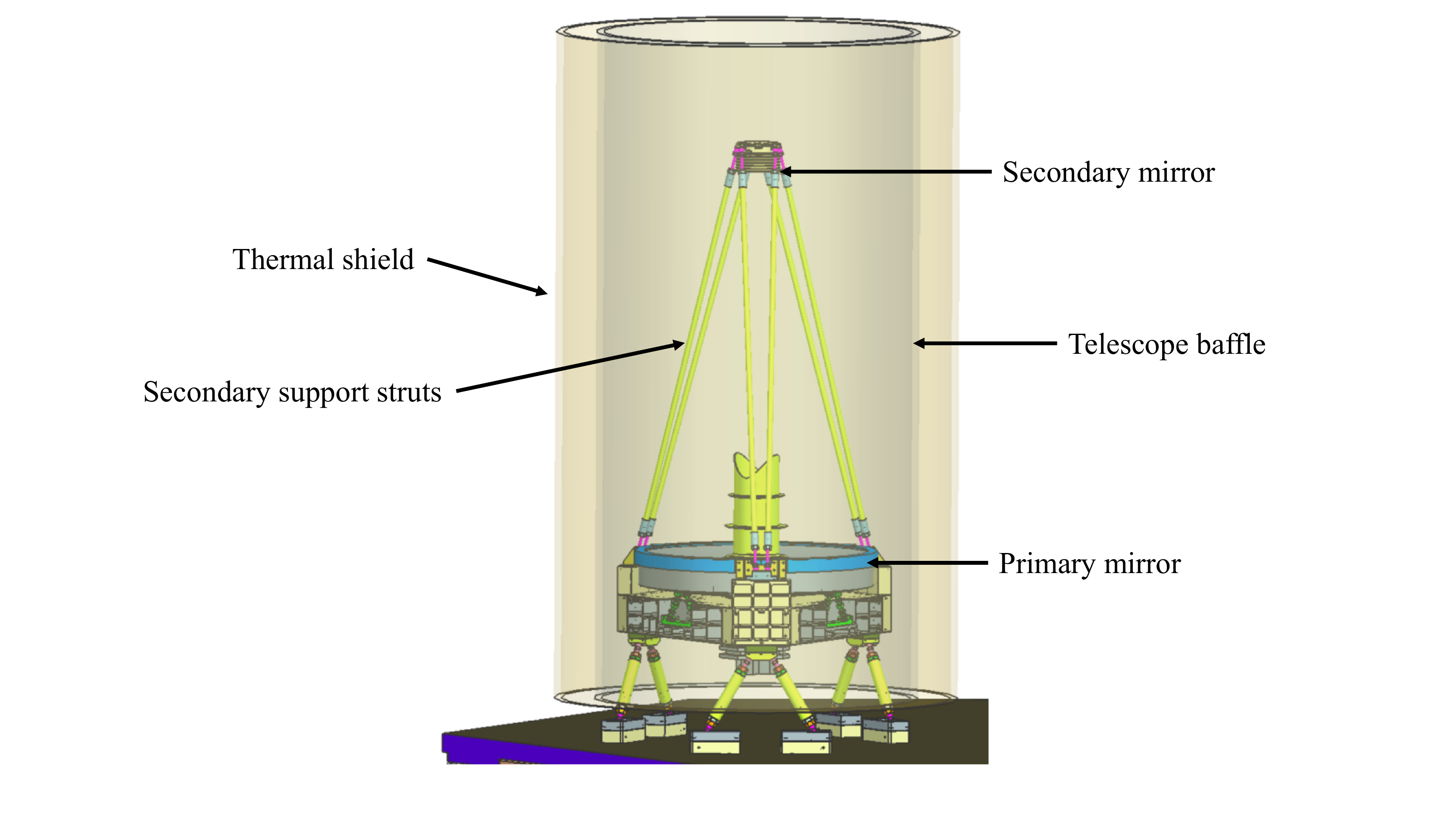}
   \end{tabular}
   \end{center}
   \caption[example] { \label{fig:PIRT} 
Overview diagram of the PIRT instrument. The detector and dichroic are located immediately below the primary mirror, but their detailed mounting structure is not yet included in our CAD model. Similarly, the instrument thermal radiators that passively cool the focal plane are not shown in this view; they will likely be mounted on the side of the telescope facing away from the payload sun shield.}
   \end{figure}

The PIRT follow-up observations consist of an initial exposure of four 125-second integrations in 5 simultaneous optical and near-infrared bands. The initial exposure and subsequent identical exposures will provide photometric measurements of potentially hundreds of astronomical objects in a 10 arcmin field of view. 
The instrument comprises a 30 cm telescope, a dichroic beam splitter, mechanical and thermal structures, a James Webb Space Telescope (JWST) flight-spare H2RG detector, a JWST flight-spare SIDECAR ASIC readout system, and instrument electronics and software.  A conceptual overview of the PIRT is depicted in Figure~\ref{fig:PIRT}. Thermal radiators, cabling, and other components are not shown, as their placement is still under development.  In the subsections below, we describe the major components of PIRT. 

%%%%%%%%%%%%%%%%%%%%%%%%%%%%%%%%%%%%%%%
%%%%%%%%%%%%%%%%%%%%%%%%%%%%%%%%%%%%%%

\subsection{Optical Assembly}
The PIRT optical assembly conceptual design consists of an optical telescope assembly (OTA) passively cooled to 200K, associated baffling to control straylight, thermal shielding, a one-time deployable aperture cover, and aft-optics. The aft-optics consists of a dichroic beam splitter, mounting structure, thermal control, and provision to mount an infrared detector and its associated cabling and readout electronics. The dichroic assembly will be passively cooled to 150K, while the detector and readout electronics will be passively cooled to 100K and 135K, respectively. The dichroic assembly additionally functions as a cold stop, limiting the detector's view of thermal emission from warm structures. The OTA is a simple, all-aluminum construction. The relatively slow beam requires no focus mechanism. 

\subsection{Dichroic Assembly}
\begin{figure} [ht]
   \begin{center}
   \begin{tabular}{c} %% tabular useful for creating an array of images 
   \includegraphics[height=5cm]{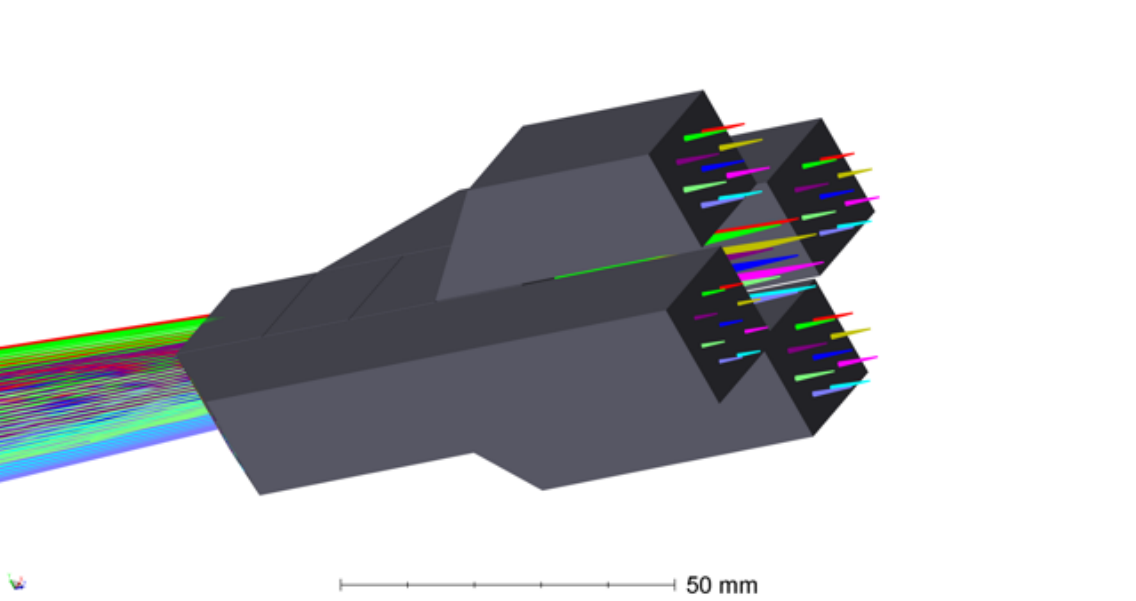}
   \end{tabular}
   \end{center}
   \caption[example] { \label{fig:dichroic1} 
Dichroic Assembly conceptual diagram. The light from the telescope enters the assembly at left and is split into 5 separate wavelength bands that arrive in simultaneous focus at the detector plane. A scale for reference appears at the bottom of the figure. }
   \end{figure}

\begin{figure} [ht]
   \begin{center}
   \begin{tabular}{c} %% tabular useful for creating an array of images 
   \includegraphics[height=5cm]{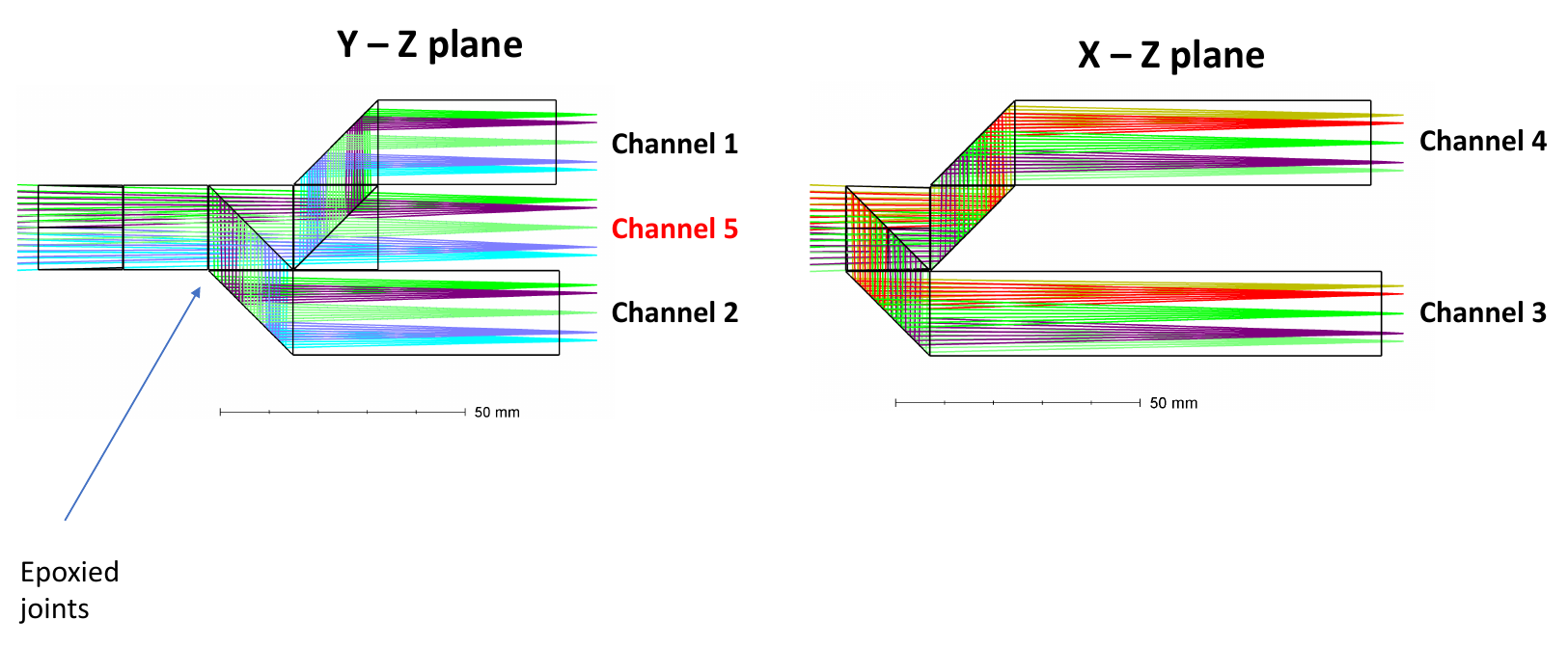}
   \end{tabular}
   \end{center}
   \caption[example] { \label{fig:dichroic2} 
Dichroic Assembly conceptual diagram. The light arriving from the telescope at left is split into 5 separate wavelength bands using a compact and simple set of glass elements bonded with optically transparent epoxy. A scale for reference appears at the bottom of the diagram. }
   \end{figure}

The purpose of the dichroic assembly is to split the light from the telescope’s field of view into 5 separate wavelength bands with the same field of view.  The optical telescope is expected to have an f-ratio of 13; the relatively slow beam allows the light from the five bands to arrive at the detector surface in common focus without the use of additional elements with optical power.  A conceptual diagram is presented in Figures~\ref{fig:dichroic1} and \ref{fig:dichroic2}. The design envisioned consists of a number of simple fused silica elements bonded with transparent epoxy. The precise definition of the band edges is still under active refinement. The nominal choice consists of five adjacent bands, each with between 25 and 35\% fractional bandwidth.

%%%%%%%%%%%%%%%%%%%%%%%%%%%%%%%%%%%%%%%
% Gamow configuration figure indicating PIRT
%%%%%%%%%%%%%%%%%%%%%%%%%%%%%%%%%%%%%%

%%%%%%%%%%%%%%%%%%%%%%%%%%%%%%%%%%%%%%%
% Gamow configuration figure indicating PIRT
%%%%%%%%%%%%%%%%%%%%%%%%%%%%%%%%%%%%%%

The requirements for the dichroic assembly performance are also under active development and have not been finalized. Two scientific goals are likely to drive the requirements: (1) total system sensitivity, and (2) the precision and freedom from systematic errors of the photometric redshift estimate.  The system sensitivity will drive requirements on the throughput of the dichroic assembly and on the overall geometric image quality. The photometric redshift objective will likely translate into requirements on the sharpness of the band edges, the allowable bandpass ripple (throughput vs wavelength within a band), and the required level of rejection of out-of-band light. We envision the requirements will be set at a level that will allow standard manufacturing techniques for the dichroic assembly and coatings, although the precise definitions are at a very early stage and await the results of further scientific modeling of the photometric redshift performance.

\subsection{Detector System}

PIRT will use Teledyne Imaging Systems Inc. (TIS) sensor, based on their 2.5 micron cutoff HgCdTe H2RG detector technology \cite{Arias1993,Loose2003} that we anticipate will be available as a JWST flight spare. This four channel, 2048x2048 pixel device will operate at 100~K and be read by a cold electronics module containing the SIDECAR application specific integrated circuit (ASIC) developed by TIS \cite{Loose2005,Loose2007}. The ASIC is an entire set of focal plane array electronics on a single chip. The cold electronics will operate at 135~K and will also available as a JWST flight spare unit. The detector will be mounted in a JWST NIRSpec flight spare mounting structure. 

\subsection{Electronics}
\begin{figure} [ht]
   \begin{center}
   \begin{tabular}{c} %% tabular useful for creating an array of images 
   \includegraphics[width=15cm]{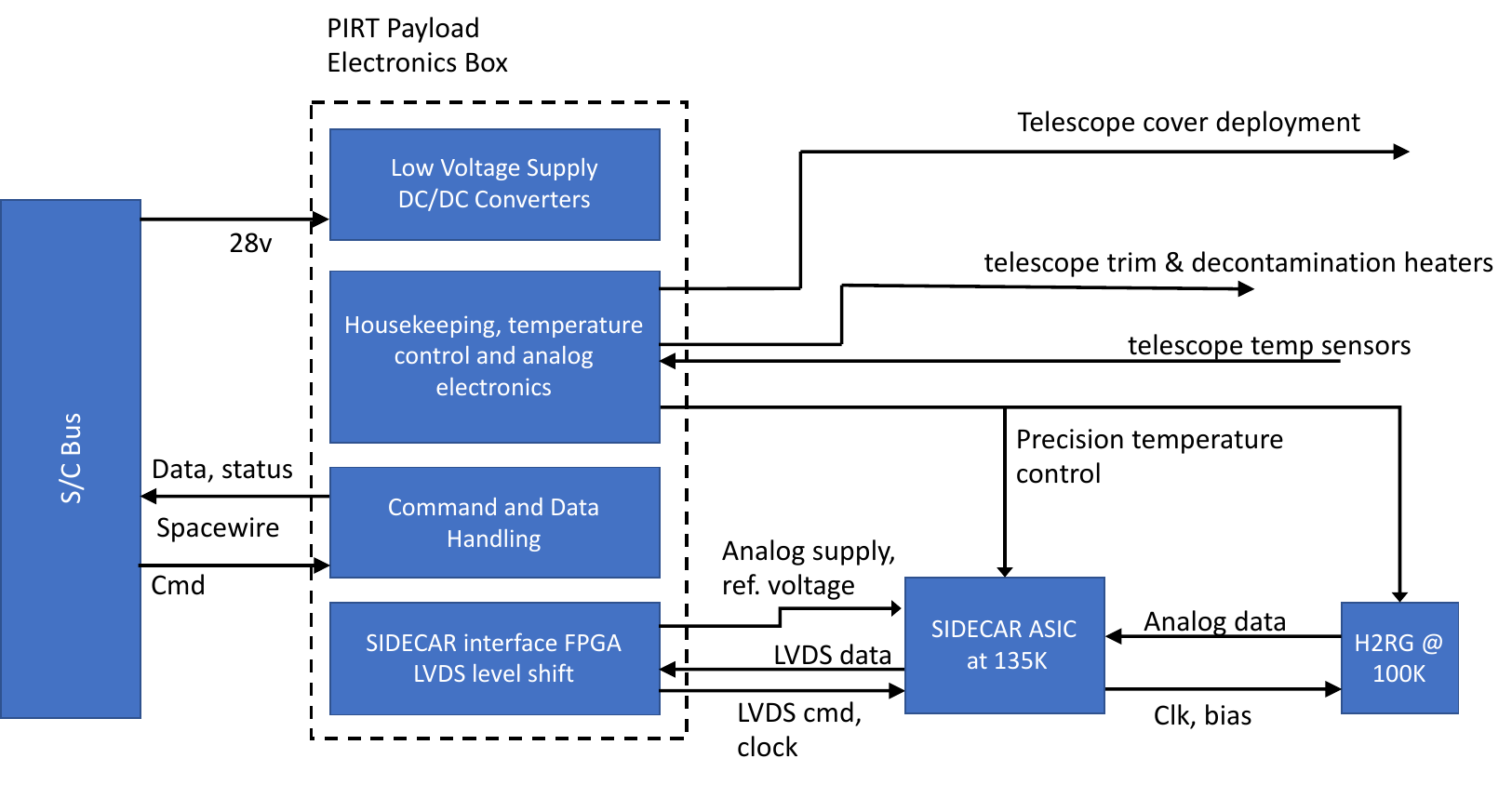}
   \end{tabular}
   \end{center}
   \caption[example] { \label{fig:electronics} 
Photo-z Infrared Telescope Electronics Conceptual Diagram }
   \end{figure}

A conceptual diagram of the PIRT instrument electronics is presented in Figure~\ref{fig:electronics}, although it should be noted that the division of functions within the electronics assembly is notional, and a final design has not been reached. The electronics functions consist of power conversion, housekeeping functions such as voltage, current, and temperature monitoring, temperature control of the detector and ASIC, temperature control of the telescope, and provision for deployment of the telescope aperture cover.  The electronics must also interface to the SIDECAR ASIC to read out the detector.  

%%%%%%%%%%%%%%%%%%%%%%%%%%%%%%%%%%%%%%%
% Gamow configuration figure indicating PIRT
%%%%%%%%%%%%%%%%%%%%%%%%%%%%%%%%%%%%%%
The detector readout scheme is expected to be sample-up-the-ramp frames at a rate of approximately one frame per 5 seconds. The JWST detector is a 4 channel device with a nominal readout time of approximately 10 seconds. The faster frame rate envisioned here would either be accomplished with a higher clock rate of 200 kHz, or with only reading a portion of the array centered around the LEXT source position. The faster frame rate allows more non-destructive reads per integration period, thus reducing the contribution of detector read noise.

The PIRT exposures are envisioned to consist of 4 integration periods of 125 seconds each. A small-angle spacecraft pointing dither will occur between integrations.  The electronics must collect the detector frames, perform reference pixel subtraction, reject cosmic rays, account for saturation, and fit the estimated per-pixel flux from the set of raw frame data.  After these calculations, the data are ready for further processing.

The overall requirements on the PIRT electronics are similar to those of the LEXT instrument in terms of input power and voltage. The principal differences are the need to provide for cover deployment; the need to provide precision temperature control of the detector and ASIC; the need to provide analog and digital interface to the ASIC, similar to other missions such as JWST, Euclid, and others; and the need to for substantial on-board processing as outlined in the following section. We are exploring the possibility of using similar designs and sub-assemblies for the PIRT and LEXT electronics.

\subsection{Software}
PIRT will combine the raw sample-up-the-ramp data frames to produce a flux estimate for each pixel in the detector array. After the low-level processing noted above, there will be additional processing split between on-board processing and on-the-ground processing.  On a weekly basis, we will have a larger available data volume for transferring the image integrations to the ground, and further processing will take place starting from that data.  For rapid follow-up of the sources with ground-based telescopes, and also with JWST, however, we will transmit data to the ground over a real-time, low data rate link. The data link does not have sufficient bandwidth to transmit the full images, so instead we will do additional processing on-board. The steps in this additional processing are still under development, but the below steps are one potential approach:
\begin{itemize}
    \item Correct or flag additional detector effects such as dead pixels and interpixel capacitance
    \item Determine the position offsets between the 4 integrations per exposure
    \item Co-add the 4 integrations per exposure after adjusting for position offsets
    \item Make a list of all sources detected at $>5 \sigma$ in the exposure in each of the 5 bands
    \item Produce a compressed list of the source positions and fluxes for transmission to the ground
\end{itemize}

The above functions drive the requirements on the capability of the on-board processing. We are still in the process of developing the detailed requirements, but our current estimate is that we will need RAM of order 128Mbytes and a processing speed comparable to a current-generation LEON4 processor.

Following the transmission of the compressed list of sources and fluxes, further processing on the ground will determine which of the sources corresponds to a new source GRB candidate and produce a photometric redshift estimate for that source. The transient source position and redshift can then be rapidly transmitted to a network of other observatories for follow-up spectroscopy. 

\section{EXPECTED PERFORMANCE}

We have modeled the performance of the PIRT instrument in the critical aspects of sensitivity, dependence of sensitivity on instrument parameters, and redshift estimation.  We describe these aspects below. By tracing the scientific performance of the instrument to its design parameters, we can rapidly evaluate how the scientific return of the mission might change with respect to different choices of instrument parameters.

\subsection{Sensitivity Model}
%%%%%%%%%%%%%%%%%%%%%%%%%%%%%%%%%%%%%%%
% Gamow configuration figure indicating PIRT
   \begin{figure} [ht]
   \begin{center}
   \begin{tabular}{c} %% tabular useful for creating an array of images 
   \includegraphics[height=11cm]{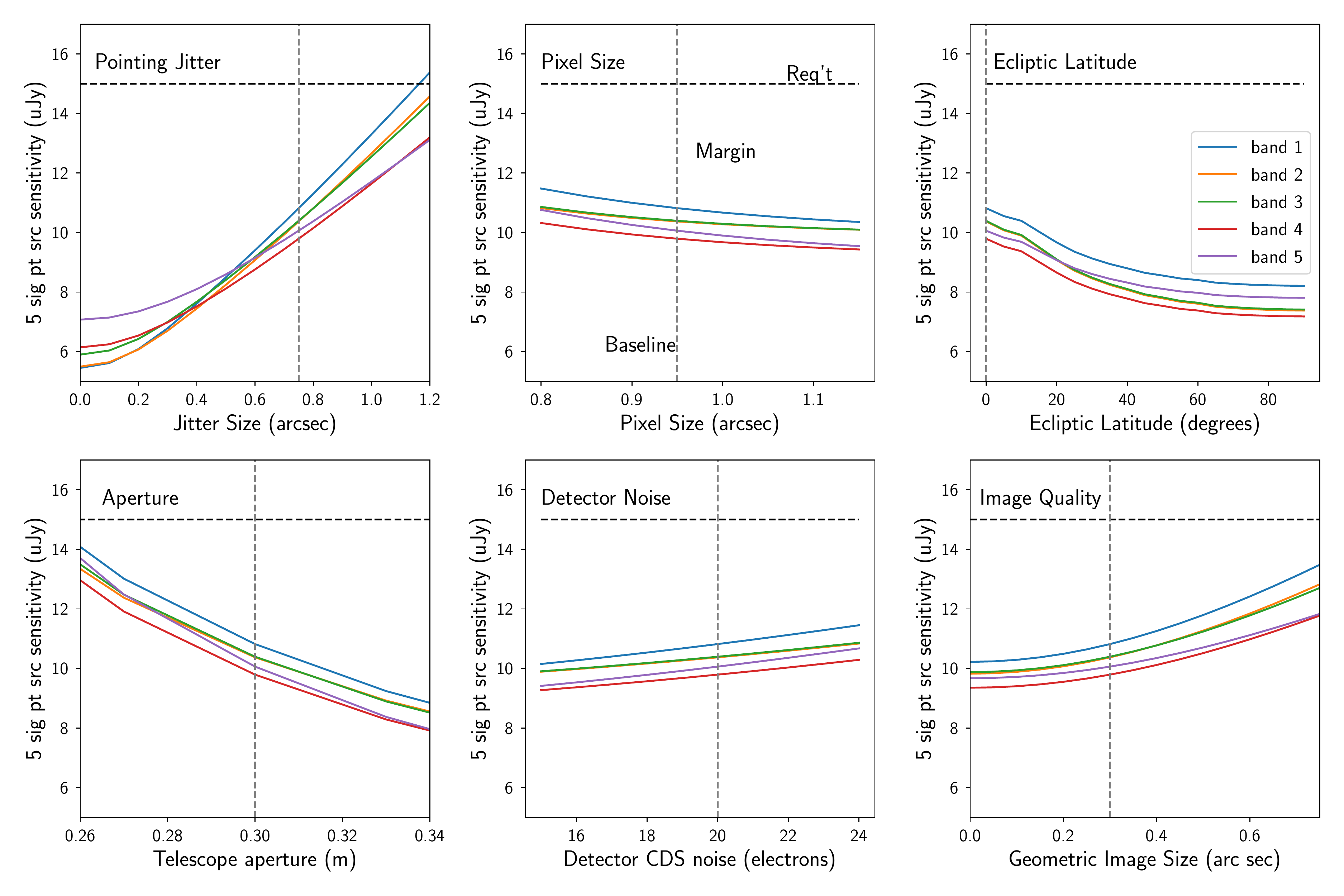}
   \end{tabular}
   \end{center}
   \caption[example] { \label{fig:dependence} 
PIRT sensitivity dependence on instrument design parameters. Each of the six panels corresponds to an example of how the PIRT 5$\sigma$ point source sensitivity varies with respect to changes in one of the instrument baseline parameters. The different colored curves represent the different PIRT wavelength bands, with band numbering increasing with wavelength. The dashed horizontal line represents the PIRT 15$\mu$Jy sensitivity requirement. The dashed vertical line represents the baseline choice of each of these instrument parameters. The vertical distance between the colored curves and the horizontal dashed line represents margin between the projected and required sensitivity.}
   \end{figure} 
%%%%%%%%%%%%%%%%%%%%%%%%%%%%%%%%%%%%%%

We model the noise equivalent flux density of the instrument using Equation~\ref{eq:sensitivity}. Here, $\Delta F_{\nu}$ is the noise equivalent flux density. The terms $T_{\rm int}$ and $t_f$ refer to the time of one detector integration and one frame time, respectively. In the equation, $\nu_0$ is the center frequency, $\Delta \nu$ is the bandwidth, $A$ is the telescope geometric area, and $\eta_{\rm TOT}$ is the total throughput of the system. The detector noise is characterized by the correlated double sample noise ($\sigma_{\rm CDS}$) and the read noise floor ($\sigma_{\rm floor}$) below which additional non-destructive reads of the detector no longer reduce the detector noise variance. The detector noise is combined with a variety of photon current noise sources: $a_1$ and $a_2$ are dimensionless constants of order unity that depend on the number of frames per integration \cite{Rauscher2007}, and 
$I_{\rm dark}$, $I_{\rm zodi}$, $I_{\rm fore}$, and $I_{\rm aft}$ are the currents associated with dark current, scattered zodiacal light, thermal emission from the telescope, and thermal emission from the aft optics, respectively.
%%%%%%%%%%%%%%%%%%%%%%%%%%%%%%%%%%
% Sensitivity Equation
\begin{equation}
\label{eq:sensitivity}
 \Delta F_{\nu}  
 = 
 \left[
 \frac{h\nu_0}{\Delta \nu T_{\rm int} \, A \, \eta_{\rm TOT}}
 \right]
 \left[{\frac{N_{\rm eff}}{N_{\rm int}}}\right]^{1/2} 
 \left[ 6 \, a_1 \,
 \left( { \frac{ t_f  }{T_{\rm int}} } \right) \, \sigma^2_{\rm CDS}
                    + \sigma^2_{\rm floor}
                     + \frac{6  }{5} \, a_2 \, T_{\rm int}\, 
                     \left(\, {I_{\rm dark} + I_{\rm zodi} + I_{\rm fore} + I_{\rm aft}} \, \right)
                     \right]^{1/2}.
\end{equation}
%%%%%%%%%%%%%%%%%%%%%%%%%%%%%%%%%%
  
PIRT is required to have a 5~$\sigma$ point source sensitivity of 15~$\mu$Jy. The sensitivity requirement is driven by the need for greater than 80\% detection of GRB afterglows. We have established a set of baseline instrument parameters that meets this requirement with margin.  Figure~\ref{fig:dependence} shows how the PIRT sensitivity varies with respect to changes in the baseline instrument parameters. In each of the figure panels, the x-axis corresponds to a variation in one of the instrument parameters, and the y-axis corresponds to the point source sensitivity. As expected, the telescope aperture and the system pointing jitter are important determinants of the sensitivity. Zodiacal light emission is an important contributor to the noise, and hence the dependence on ecliptic latitude.

\subsection{Redshift estimate}

\begin{figure}
    \centering
    \includegraphics[width = 0.44\linewidth]{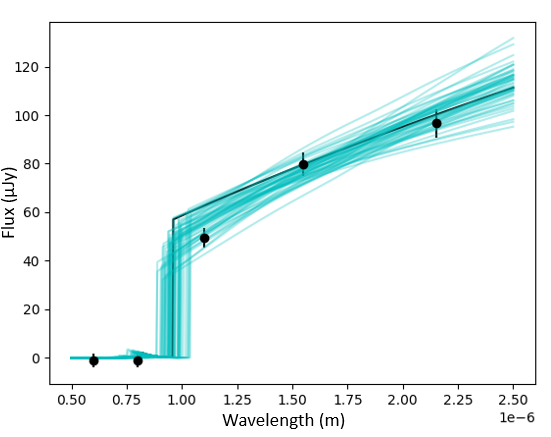}
    \includegraphics[width = 0.55\linewidth]{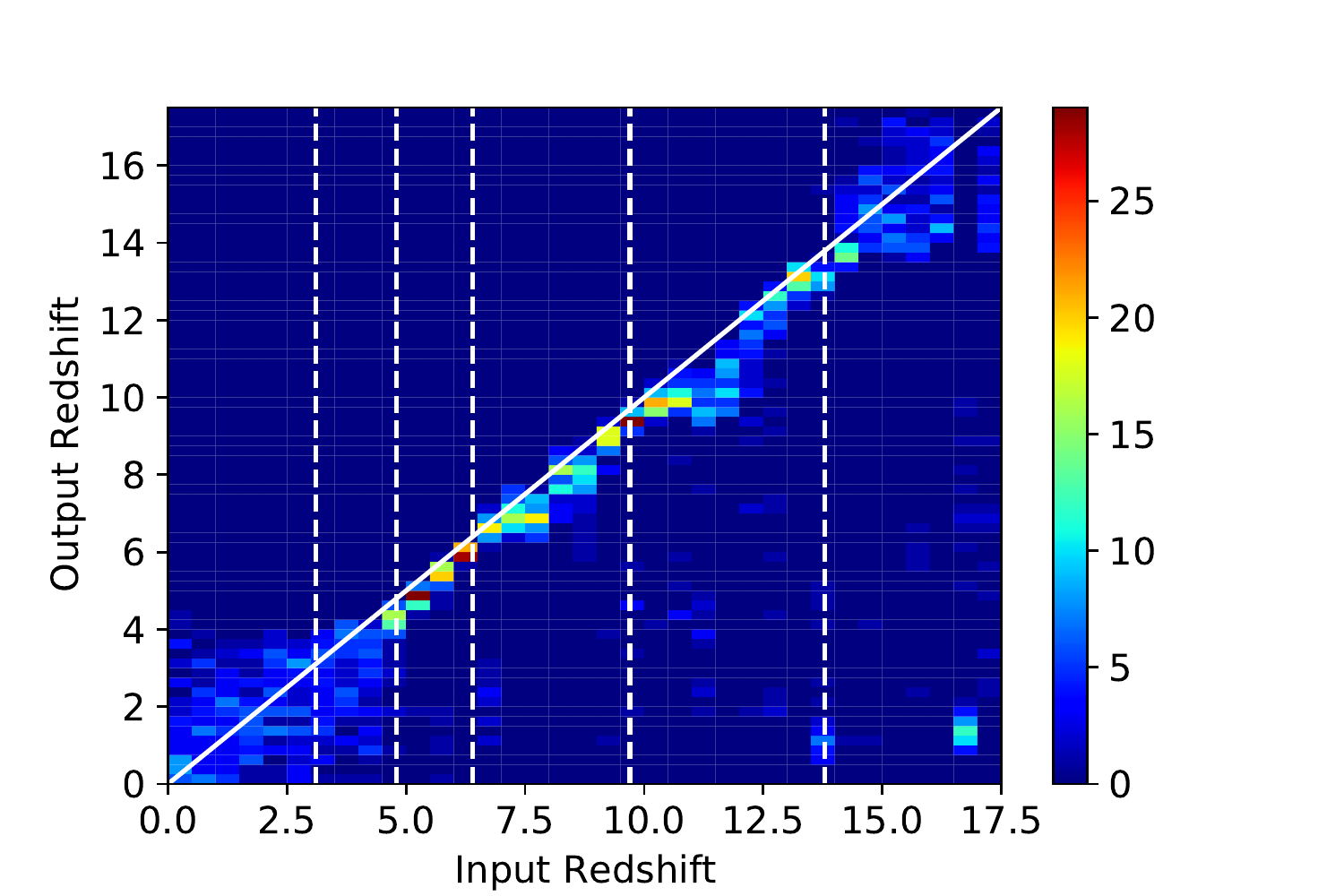}
    \caption{{\bf Left:} Fitting results for a simulated gamma-ray burst at a redshift of 7. The black line is the original spectrum, the data points are the perturbed fluxes in each filter with error bars, and the colored curves are the results for each of the walkers in the MCMC fits. The sharp drop-off in the spectrum results from the Lyman-$\alpha$ forest. 
    {\bf Right:} Density plot showing the input versus recovered/output redshift for our fitting method. The solid white shows where the input redshift is the same as the output redshift, and the dashed white lines show where the Lyman-$\alpha$ line corresponds to the filter edges.}
    \label{fig:photozplot}
\end{figure} 

We are performing simulations to assess and optimize the PIRT for determining photometric redshifts. 
These simulations assume a simple power-law spectrum for the gamma-ray burst emission in the near-infrared regime, consistent with theoretical expectations and observations of gamma-ray bursts. 
In the simulations, we apply two main effects on the emitted spectrum: intergalactic attenuation of the light \cite{meiksin} and extinction in the host galaxy \cite{pei}. 
Given parameters for the input spectrum and the light attenuation models, we simulate flux measurements for each of the PIRT's 5 photometric bands. 
In particular the Lyman-$\alpha$ forest, as part of the intergalactic attenuation, causes a sharp drop-off in the measured GRB spectrum that can be used for determining the redshift. 
The flux values for each band are perturbed by drawing from a Gaussian distribution with 5\% systematic uncertainty (to account for calibration errors) added in quadrature to 3~$\mu$Jy instrumental noise. 

One possible and promising approach to fitting the spectra is the Markov-Chain Monte-Carlo (MCMC) method, using the \texttt{emcee} package \cite{emcee}.
The simulated spectra are fit using this method to determine the best fit parameters for the amplitude, spectral index, redshift, and host galaxy extinction. 
The fitting method includes priors for the spectral index and extinction, based on large gamma-ray burst studies \cite{covino}. An example simulated spectrum is shown on the left-hand side of Figure~\ref{fig:photozplot}, while the right-hand side of this figure shows preliminary results for input redshift and recovered redshift for a certain set of parameters.

\section{CONCLUSION}

We have described the PIRT instrument, a powerful yet simple instrument for localizing GRBs and estimating their redshift. Aside from a one-time deployed cover, the instrument has no moving parts, relies on passive cooling, and uses an all-aluminum telescope design. The instrument concept is being developed for an exciting mission, the Gamow Explorer, currently expected to be proposed in 2021 for a NASA MIDEX astrophysics mission. 

\acknowledgments % equivalent to \section*{ACKNOWLEDGMENTS}       
It is a pleasure to thank the extended Gamow team for their work in defining the mission proposal.  The research described in this paper was performed in part at the Jet Propulsion Laboratory, California Institute of Technology, under a contract with the National Aeronautics and Space Administration. © 2021. All rights reserved.

% References
\bibliography{report} % bibliography data in report.bib
\bibliographystyle{spiebib} % makes bibtex use spiebib.bst

\end{document}